# Optical guided dispersions and subwavelength transmissions in dispersive plasmonic circular holes

Ki Young Kim[1], Young Ki Cho[1], Heung-Sik Tae[1], and Jeong-Hae Lee[2]

[1]School of Electrical Engineering and Computer Science, Kyungpook National University, Daegu 702-701, Korea,

[2]Department of Radio Science and Communication Engineering, Hongik University, Seoul 121-791, Korea

The light transmission through a dispersive plasmonic circular hole is numerically investigated with an emphasis on its subwavelength guidance. For a better understanding of the effect of the hole diameter on the guided dispersion characteristics, the guided modes, including both the surface plasmon polariton mode and the circular waveguide mode, are studied for several hole diameters, especially when the metal cladding has a plasmonic frequency dependency. A brief comparison is also made with the guided dispersion characteristics of a dispersive plasmonic gap [K. Y. Kim, *et al.*, Opt. Express **14**, 320-330 (2006)], which is a planar version of the present structure, and a circular waveguide with perfect electric conductor cladding. Finally, the modal behavior of the first three TM-like principal modes with varied hole diameters is examined for the same operating mode.

**Keywords:** dispersion, dispersive plasmonic hole, subwavelength guidance, surface plasmon polariton, surface wave.

## 1. Introduction

At optical frequencies, circular dielectric waveguides with metal cladding have been extensively utilized as probes for near-field scanning optical microscopy (NSOM) [1, 2]. To obtain better resolutions for NSOM systems, the diameters of the probe tips are usually reduced to dimensions much smaller than the operating wavelengths of the incident light. Yet, conventional thinking assumes that a probe tip with a small aperture has a bad throughput property [3] – the power throughput is reduced to a quarter of the power of the aperture dimensions. However, this limitation can be overcome by exciting or coupling the surface plasmon polaritons (SPPs) [2, 4, 5] due to the negative permittivity of the metal cladding. Such a scheme allows the subwavelength dimensions of the probes to be widely used with a high throughput in current cutting-edge technologies, such as biological imaging and probing [1], single molecule studies [1], and microelectronic circuit characterization and failure analysis [6]. Thus, understanding the fundamental light propagation along the probe structure, *i.e.*, a circular dielectric waveguide with metal cladding, is very important for the effective design of probes that play a critical role in improving the overall performance of an NSOM system.

Another recent finding associated with circular dielectric waveguides with metal cladding is enhanced light transmission phenomena through a single subwavelength circular hole in a thick conducting screen [7-9], which is also contrary to classical notions [3]. The light transmission process through a subwavelength circular hole with a finite thickness also includes the existence of propagating modes down to the guiding structure using the finite thickness of the conductors between both aperture exits, where the propagating modes and free space modes couple and decouple with each other. Thus, examining the subwavelength propagating modes guided along the structure is also crucial before a further detailed analysis of the manifestations of the enhanced light transmission.

Enhanced transmission has already been reported through subwavelength apertures on metal screens, including subwavelength guiding structure supporting TEM modes, *e.g.*, subwavelength coaxial apertures [10, 11] and subwavelength thick slits [12, 13]. The dimensions of the waveguide cross-section can be reduced to much less than the wavelengths due to the absence of a cutoff wavelength for the TEM modes. In contrast, it is well known that metal waveguides with a circular cross-section have cutoff wavelengths, *i.e.*, the mode cannot propagate below a certain critical diameter, if the metal is considered to be a perfect electric conductor (PEC) [14]. Practically, at optical frequencies, metals are no longer PECs, as the real parts of the permittivity become negative. As such, an SPP mode propagation can be propagated even when the diameter of the hole is much smaller than the guided wavelength. Such SPP mode propagations have already been investigated along circular dielectric waveguides with plasmonic cladding [15-19]. Furthermore, waveguide modes with modal properties similar to those of circular PEC waveguides are also known to exist together with SPPs [20, 21]. The possibility of electromagnetic energy transfers along a long circular air channel with "real" plasma cladding has also been proposed [22, 23], where the propagation mechanisms are in principle the same as those of the plasmonic response of a hollow circular waveguide with metal cladding. Quite recently, the effects of the plasmonic dispersion relation on the transmission properties of subwavelength cylindrical holes have been considered [24]. However, despite extensive research, the guided dispersion characteristics of circular dielectric waveguides with plasmonic cladding still need to be investigated further. More specifically, the effect of the diameter of a circular waveguide with dispersive plasmonic cladding on the dispersion characteristics has not been completely investigated. Thus, the guided dispersion characteristics along a subwavelength circular hole with plasmonic cladding need to be investigated.

Accordingly, this paper theoretically examines the guided dispersion characteristics of a circular air channel surrounded by metal using its plasmonic permittivity response, *i.e.* the Drude model. In particular, the role of the subwavelength-hole diameter is taken into account. Some new findings are discussed and compared with the characteristics of other relevant guiding structures, such as dispersive plasmonic gaps (DPGs) [25].

## 2. Dispersive plasmonic hole (DPH): Structure

Figure 1 shows a schematic illustration of the dispersive plasmonic hole (DPH) under consideration, which is a circular air channel (region 1, $r < a$) surrounded by perforated bulk metal (region 2, $r > a$). The dielectric constant of the air core and magnetic constants (relative permeability) of both regions are identically assumed to be $\varepsilon_{r1} = \mu_{r1} = \mu_{r2} = 1.0$. The dielectric constants of metals at optical frequencies can generally be fitted to the Drude model as follows:

$$\varepsilon_r(\omega) = 1 - \frac{\omega_p^2}{\omega(\omega - j\omega_\tau)} \tag{1}$$

where $\omega_p$ and $\omega_\tau$ are the plasma and collision frequencies, respectively. Here, we chose $f_p = \omega_p / 2\pi = 3600$ THz, which is similar to that for titanium (Ti) and already widely utilized in other calculations [25-35]. The collision frequency is given as $\omega_\tau / 2\pi = 340$ THz [26-35], which is slightly smaller than the plasma frequency, typically $\omega_\tau < 0.1\omega_p$. Practically, since a small collision frequency value has a minimal effect on the dispersion characteristics, the effect of $\omega_\tau$ is disregarded for simplicity, *i.e.*, $\omega_\tau = 0.0$. Another benefit to choosing a real dielectric constant is to point out the distinction between the two operating modes, the SPP and circular waveguide (CWG) modes, as will be discussed later. An additional Lorentzian resonance effect should be included in (1), if the operating frequencies are much lower than the plasma frequency [20, 36]. However, in the present study, eigenmode solutions become available with relatively higher frequencies, *i.e.* near and above 1000 THz, as discussed in section 4. As such, an approximate expression is used for the dielectric constant of the plasmonic cladding, *i.e.*, $\varepsilon_{r2} = 1 - \omega_p^2 / \omega^2$ with $\omega_p / 2\pi = 3600$ THz, as illustrated in Fig. 2. As expected, negative and positive values for the dielectric constants occurred below and above the plasma frequency, respectively. Plus, there was a critical frequency of $f_c = f_p / \sqrt{2} = 2545.58$ THz, at which the dielectric constant was negative unity. The significance of the critical frequency on the guided dispersion characteristics will be described in section 4. Below the plasma frequency, the refractive index of the cladding $n_2 = (\varepsilon_{r2}\mu_{r2})^{1/2}$ was purely imaginary due to the negative value of the dielectric constant. In contrast, above the plasma frequency, a positive real (but less than unity) refractive index $n_2 = (\varepsilon_{r2}\mu_{r2})^{1/2}$ was obtained, as in the case of conventional dielectric media, and plotted as a dashed line in Fig. 2.

## 3. Dispersive plasmonic hole (DPH): Characteristic equations

The characteristic equations for a DPH [24] are literally identical to those for plasma or metal cylinders embedded in an infinite surrounding dielectric medium [37] if general Bessel and Hankel functions are used to describe the fields [38]. However, the characteristic equations for DPHs can be treated in more detail if purely guided modes without any loss, including radiation, are considered. As such, the characteristic equations result from enforcing the appropriate boundary conditions. The tangential field components are divided into the axial components, as given in Table 1, and the azimuthal field components, which can readily be obtained from the axial components [14]. The characteristic equations for DPHs can be given as $F_1 F_2 = G^2$, where $F_1$, $F_2$, and $G$ are the primary characteristic functions. The details of the primary characteristic functions and other associated belongings are given in Table 1, where $k_i (i = 1, 2)$ is the transverse propagation constant in each region, $k_0$ is the free space wave number, and $\bar{\beta}(= \beta / k_0)$ is the normalized propagation constant, which is the propagation constant normalized by the number of free space waves.

In the present case, two operating modes were considered: the SPP mode and CWG mode. Detailed descriptions of both operating modes are discussed with numerical examples in section 4. $J_m(\cdot)$ and $I_m(\cdot)$ are ordinary Bessel and modified Bessel functions of the first kind and describe the fields inside the boundary ($r < a$) of each mode, respectively, where $m$ is the azimuthal eigenvalue. In both cases, the fields along the cladding region ($r > a$) are governed by a normalized Bessel function of the second kind, *i.e.*, $K_m(\cdot)$, to satisfy the radiation condition at infinity. In the negative permittivity region, *i.e.*, below the plasma frequency, the conditions for the SPP and CWG modes are $\bar{\beta} > (\mu_{r1}\varepsilon_{r1})^{1/2}$ and $\bar{\beta} < (\mu_{r1}\varepsilon_{r1})^{1/2}$, respectively. However, above the plasma frequency, the conditions for the CWG mode become $(\mu_{r1}\varepsilon_{r1})^{1/2} > \bar{\beta} > (\mu_{r2}\varepsilon_{r2})^{1/2}$ due to the real value for the refractive index of the cladding, *i.e.*, $n_2 = (\mu_{r2}\varepsilon_{r2})^{1/2}$, and the SPP mode cannot exist. When there is no azimuthal variation, *i.e.*, $m = 0$, the characteristic equation is split into $F_1 = 0$ and $F_2 = 0$ for the $TM_{0n}$ and $TE_{0n}$ modes, respectively. Meanwhile, if there is an azimuthal variation, *i.e.*, $m \geq 1$, the characteristic equation $F_1 F_2 = G^2$ can be rewritten using an empirical induction procedure to distinguish the $HE_{mn}$ and $EH_{mn}$ modes as follows:

$$\left(\frac{\mu_{r2}}{\mu_{r1}} + \frac{\varepsilon_{r2}}{\varepsilon_{r1}}\right)\frac{Q}{2} \pm \left[\left\{\left(\frac{\mu_{r2}}{\mu_{r1}} + \frac{\varepsilon_{r2}}{\varepsilon_{r1}}\right)\frac{Q}{2}\right\}^2 + \frac{R}{\mu_{r1}\varepsilon_{r1}}\right]^{1/2} - P = 0. \tag{2}$$

Here, the secondary characteristic functions $P$, $Q$, and $R$ are also given in Table 1. The "$\pm$" signs in (2) are for the $HE_{mn}$ and $EH_{mn}$ modes, respectively, which have similar dispersion characteristics to the $TM_{0n}$ and $TE_{0n}$ modes, respectively. Thus, for convenience, the $TM_{0n}$ and $HE_{mn}$ modes are called the TM-like modes, while the $TE_{0n}$ and $EH_{mn}$ modes are called the TE-like modes. The characteristic equations are numerically solved and numerical illustrations with the normalized propagation constants versus the operating frequencies shown in the next section.

## 4. Numerical results and discussion

### 4.1. Dispersion characteristics depending on hole diameter

Figure 3 shows the dispersion characteristics for the Ti-like plasmonic medium cladded DPHs with various hole diameters of $D = 150, 100,$ 70, 40, 10, and 2 nm. For the lowest three azimuthal eigenvalues $(m = 0,\ 1,\ 2)$, the normalized propagation constants were representatively obtained, as higher order hybrid modes with $m \geq 3$ are believed to have similar dispersion characteristics to the cases of $m = 1$ and 2. The shaded regions in Fig. 3 are the forbidden regions, *i.e.*, $\bar{\beta} < (\mu_{r2}\varepsilon_{r2})^{1/2}$, which occurred above the plasma frequency of $f_p = 3600$ THz. The solutions for the higher order modes of the TM-like modes and all the TE-like modes remained in the fast wave region, *i.e.*, CWG mode region. CWG modes existed both below and above the plasma frequency. However, the sequence of successive higher order mode generation was limited above the plasma frequency. The $EH_{21}$ mode for the $D = 100$ nm case in Fig. 3(b) and $EH_{11}$ mode for the $D = 40$ nm case in Fig. 3(d) had their cutoff on the line $n_2 = (\mu_{r2}\varepsilon_{r2})^{1/2}$. Many higher order modes are seen in Fig. 3(a), which were obtained in the case of relatively larger hole diameter. The higher order modes of the TM-like modes and TE-like modes were observed to be suppressed as the hole diameter decreased. Yet, it is interesting to note that, even at very small subwavelength diameters, the three principal modes ( $n = 1$ ) among the TM-like modes, *i.e.*, the $TM_{01}$, $HE_{11}$, and $HE_{21}$ modes, still remained without being suppressed, as shown in Fig. 3(e) and (f). Thus, investigating the dispersion characteristics of the TM-like modes became the main focus, as they were able to propagate regardless of how small the hole diameter was and exhibit subwavelength guidance. Only the principal TM-like modes had guided mode solutions in the slow wave region ( $\beta / k_0 > 1.0$ ), corresponding to the SPP mode region. The normalized propagation constants increased asymptotically to infinity as the frequency approached the critical frequency of $f_c = 2545.58$ THz.

Otherwise, the dispersion curves continued smoothly to the fast wave region and were finally cutoff ( $\beta / k_0 = 0$ ). The cutoff frequencies for the $TM_{01}$, $HE_{11}$, and $HE_{21}$ modes increased as the hole diameters decreased. With a larger hole diameter, as shown in Fig. 3(a), the sequence of the modes was $HE_{11}$, $TM_{01}$, and $HE_{21}$, which was also previously observed by Novotny *et al.* [20], whereas for smaller diameters, the sequence of $HE_{11}$, $HE_{21}$, and $TM_{01}$ was noted. This observation was due to the material dispersion of the plasmonic cladding based on its frequency dependent nature. The cutoff frequencies for the $HE_{11}$ mode were always lower than those for the $HE_{21}$ mode, as shown in Fig. 3. Meanwhile, the cutoff frequency for the $TM_{01}$ mode in the case of Fig. 3(a) was found to be between those for the $HE_{11}$ and $HE_{21}$ modes. However, in the cases of Fig. 3(b)-(f) with smaller hole diameters, the cutoff frequencies for the $TM_{01}$ mode were higher than those for the other modes. As the diameter of the hole decreased, the cutoff frequency for the $TM_{01}$ mode increased towards the plasma frequency $\left(f_p = 3600 \text{ THz}\right)$, while cutoff frequencies for the other modes approached the critical frequency $\left(f_c = 2545.58 \text{ THz}\right)$. With very small subwavelength diameters, the propagation of the $TM_{01}$ mode was a backward wave (corresponding to the negative slopes in the dispersion curves), which only existed in the frequency region between the critical and plasma frequencies. Also, forward waves (corresponding to the positive slopes in the dispersion curves) were only available with the hybrid modes near the critical frequency, which is inconsistent with previous results for DPGs [25], where eigenmode solutions for the guided modes have been found to exist throughout the frequency range. The SPPs propagated below the critical frequency, and the waveguide modes also existed above the plasma frequency even with a very narrow subwavelength gap width. However, the guided dispersion characteristics of the $TM_{01}$ mode (for the DPH) and $TM_1$ mode (for the DPG) were quite similar (essentially the same) to each other due to the plasmonic response of the metal cladding.

### 4.2. Relation with circular PEC waveguides

Since metals are usually modeled as PECs, the guided dispersion characteristics of a DPH need to be compared with those of a circular PEC waveguide, which is the idealized version of a DPH. The normalized propagation constant of circular PEC waveguides for the $TE_{mn}$ and $TM_{mn}$ modes are given respectively as follows [14]:

$$\bar{\beta} = \frac{\beta}{k_0} = \left\{1.0 - \left(\frac{p'_{mn}}{k_0 a}\right)^2\right\}^{1/2} \quad \text{for } TE_{mn} \text{ mode} \qquad (3a)$$

$$\bar{\beta} = \frac{\beta}{k_0} = \left\{ 1.0 - \left( \frac{p_{mn}}{k_0 a} \right)^2 \right\}^{1/2} \quad \text{for } \mathrm{TM}_{mn} \text{ mode,} \tag{3b}$$

where $p'_{01} = 3.832$, $p'_{11} = 1.841$, $p'_{21} = 3.054$, $p_{01} = 2.405$, $p_{11} = 3.832$, and $p_{21} = 5.135$. It should be noted that the values of $p'_{01}$ and $p_{11}$ were the same, that is, $p'_{01} = p_{11} = 3.832$. This means, as expected, the $TE_{01}$ and $TM_{11}$ modes for the circular PEC waveguide degenerated. Figure 4 shows the relation between the DPH and circular PEC waveguides on the dispersion curves for the case $D = 150$ nm. It is well known that the cutoff frequencies of plasmonic waveguides are lower than those of PEC waveguides with the same geometry [24, 25, 39]. This property due to the penetration of the evanescent fields into the plasmonic medium was also observed with the present structure.

Furthermore, the relationship of the guided dispersions between circular waveguides with a PEC and the plasmonic cladding was also clarified. The $HE_{m1}$ ($EH_{m1}$) modes ($m \geq 1$) correspond to the $TE_{m1}$ ($TM_{m1}$) modes [24, 40], while the $TM_{01}$ and $TE_{01}$ modes for both cases correspond to the modes with the same designations. As the azimuthal eigenvalue increased, the difference between the cutoff frequencies for the DPH and circular PEC waveguides also increased. The TE-like modes such as $TE_{01}$, $EH_{11}$, and $EH_{21}$ modes all exist in the CWG mode regions. As shown in Fig. 4(b), the $EH_{11}$ and $TE_{01}$ modes were quite similar, yet did not degenerate any further. This deviation was also due to the plasmonic response of the metal cladding for the different polarizations. For smaller diameters, *e.g.*, subwavelength diameters, more conspicuous differences were observed between the dispersion curves for the DPH and circular PEC waveguides, possibly because more power was propagated along the cladding region of the DPH, whereas such a phenomenon could not occur in the PEC counterpart.

### 4.3. Dispersion characteristics of TM-like principal modes

As shown in Fig. 3, the principal modes of the TM-like modes, *i.e.*, the $TM_{01}$, $HE_{11}$, and $HE_{21}$ modes, supported the SPP modes, which play significant roles in subwavelength transmission. These modes including subwavelength SPP modes might be excited with some tapered structures described in ref. [4], for instance. Here, the evolution of these modes relative to the hole diameter was investigated, and more distinctive features obtained based on plotting the modes with varied hole diameters as a parameter for the same operating mode. Figure 5 shows the dispersion curves for the principal TM-like modes of the first (lowest) three azimuthal eigenvalues, the same data as in Fig. 3. In Fig. 5(a), the cutoff frequencies for the $TM_{01}$ modes increased up to the plasma frequency as the hole diameter decreased. For smaller diameters, *i.e.*, $D = 40$, 10, and 2 nm, the propagations were all backward waves. For the hybrid modes, the dispersion curves for the smaller diameters also had bifurcation points, where forward and backward waves were generated simultaneously, as shown by the insets in Fig. 5(b) and (c). In particular, eigenmode solutions for the hybrid modes with small diameters, *e.g.*, $D = 10$, 2 nm, were only available near the critical frequency, *i.e.*, $f_c = 2545.58$ THz, allowing a wide single mode operation region for the backward $TM_{01}$ mode. The backward waves along these nonperiodic waveguiding structures might be utilized in wide application areas of mode filters, switching, phase compensating, frequency selections, and many others, as suggested in ref. [41, 42].

For practical applications of the DPH, the attenuation characteristics should be taken into account, which is, however, beyond the scope of this work. The propagation length (or path length) of the DPH will be limited by the nonzero value of the collision frequency ($\omega_\tau$) in (3) and some other relevant parameters [18]. As the hole diameter decreases, the (subwavelength) modes might experience higher attenuations because more electromagnetic fields penetrate into the cladding region in evanescent forms. Also, their attenuation behaviors with dissimilar azimuthal eigenvalues are known to be different with each other [18, 20] due to the different field configurations.

## 5. Conclusions

The light transmission along a DPH was numerically investigated and discussed. Especially, the hole-diameter dependency of the dispersion characteristics was analyzed with an emphasis on the subwavelength guidance. The results presented here were also compared with those for a dispersive plasmonic gap, which is a planar version of a DPH. Only the principal modes of the TM-like modes existed in the slow wave region, where the SPP modes were supportable. The higher order mode of the TM-like modes and all the TE-like modes were observed to be guided in the fast wave regions, corresponding to the CWG mode. A brief comparison of DPH and circular PEC waveguides revealed that the cutoff frequency for the DPH shifted toward the lower frequency region, which was even more conspicuous for higher azimuthal eigenvalues. Under very small subwavelength-hole conditions, only the principal modes of the TM-like mode were propagated. The $TM_{01}$ mode as a backward wave was observed to be guidable over the frequency range from the critical frequency to the plasma frequency. Higher order modes, such as the $HE_{11}$ and $HE_{21}$ modes, were found to exist within a very narrow frequency region near the critical frequency. Although the plasmonic medium used in this investigation was restricted to a Ti-like medium, the present results can still be applied to other DPHs with different metal claddings that have plasmonic responses. Consequently, the present results can provide some guidelines for the design of optical devices based on the dispersion characteristics of a DPH.

**Acknowledgment**

This work was supported by grant No. R01-2004-000-10158-0 from the Basic Research Program of the Korea Science & Engineering Foundation.

## Figure Captions

**Fig. 1.** Schematic illustration of dispersive plasmonic circular hole. The hollow core and infinite cladding are the free space and metal, respectively. The dielectric constant of the metal can be fitted to the Drude model.

**Fig. 2.** Dielectric constant for Ti-like metal cladding. Below the plasma frequency, the relative permittivity is negative, and there is a critical frequency of $f_c = 2545.58$ THz at which the dielectric constant for the cladding is negative unity. Then, above the plasma frequency of $f_p = 3600$ THz, the dielectric constant is positive, at which point the refractive index becomes a positive real value, which is plotted by a dashed line.

**Fig. 3.** Dispersion curves for DPHs in case of (a) $D = 150$ nm, (b) $D = 100$ nm, (c) $D = 70$ nm, (d) $D = 40$ nm, (e) $D = 10$ nm, and (f) $D = 20$ nm. The shaded region delimited by the dotted lines is the forbidden region with no propagating mode.

**Fig. 4.** Relation between DPH and circular PEC waveguide based on dispersion curves. In both cases, the diameter is $D = 150$ nm. (a) TM-like mode for DPH and PEC correspondents, and (b) TE-like modes for DPH and PEC correspondents. The cutoff frequency for the DPH is downshifted when compared with the PEC correspondents.

**Fig. 5.** Dispersion curves for principal TM-like modes of first three azimuthal eigenvalues, *i.e.*, (a) $TM_{01}$, (b) $HE_{11}$, and (c) $HE_{21}$ modes. The shaded regions delimited by the dotted lines are forbidden regions. The circles in the insets represent the bifurcation points where the forward and backward waves branch off.

**Figure 1**

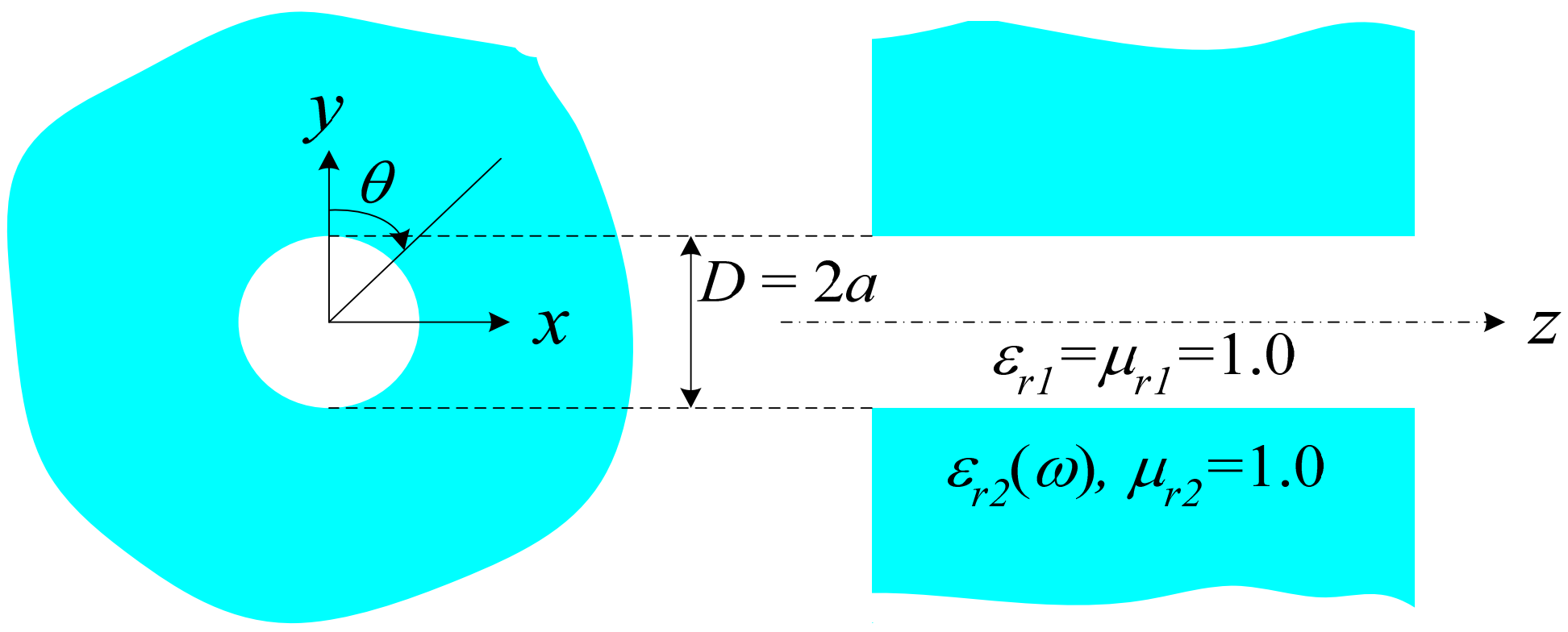

**Figure 2**

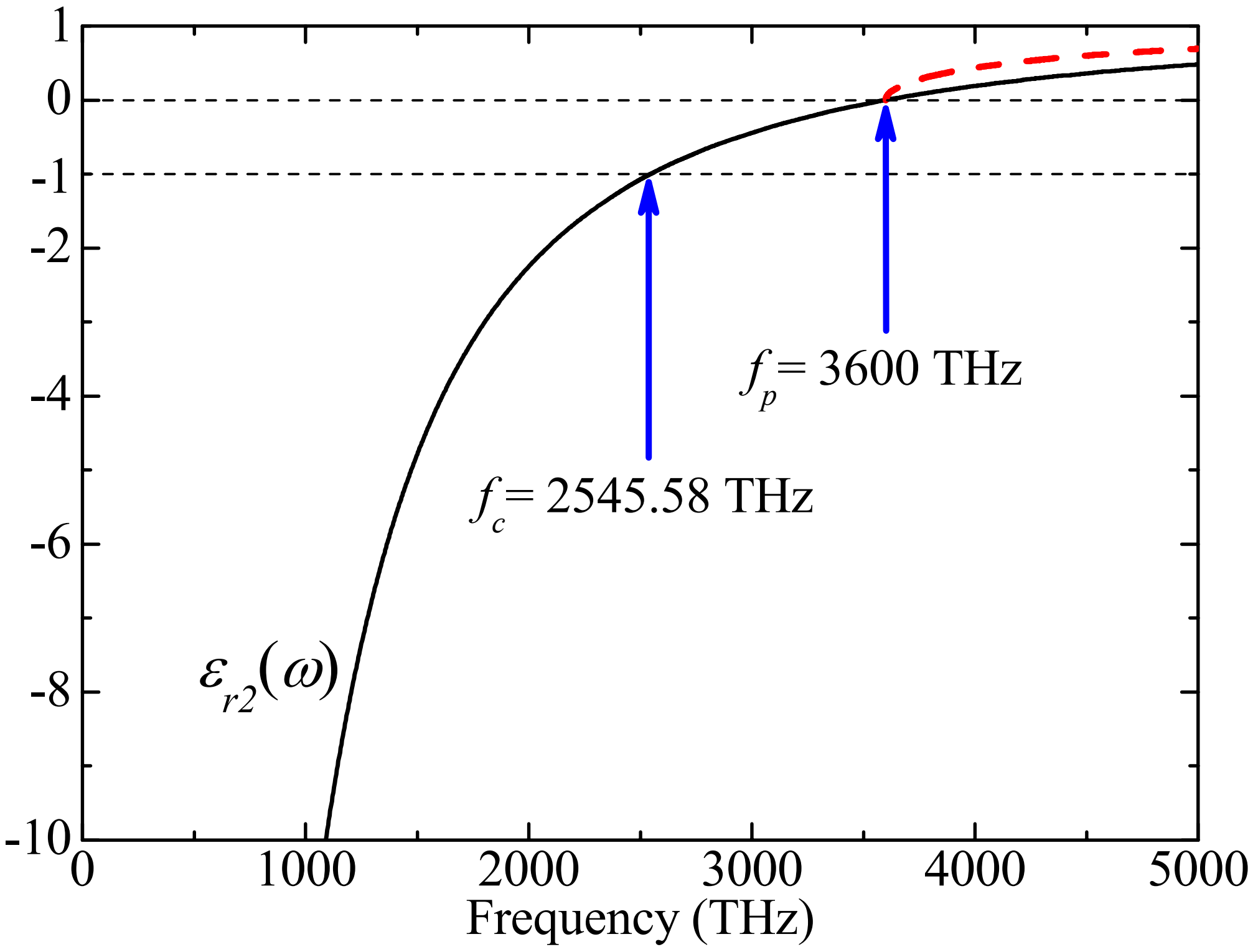

# Figure 3

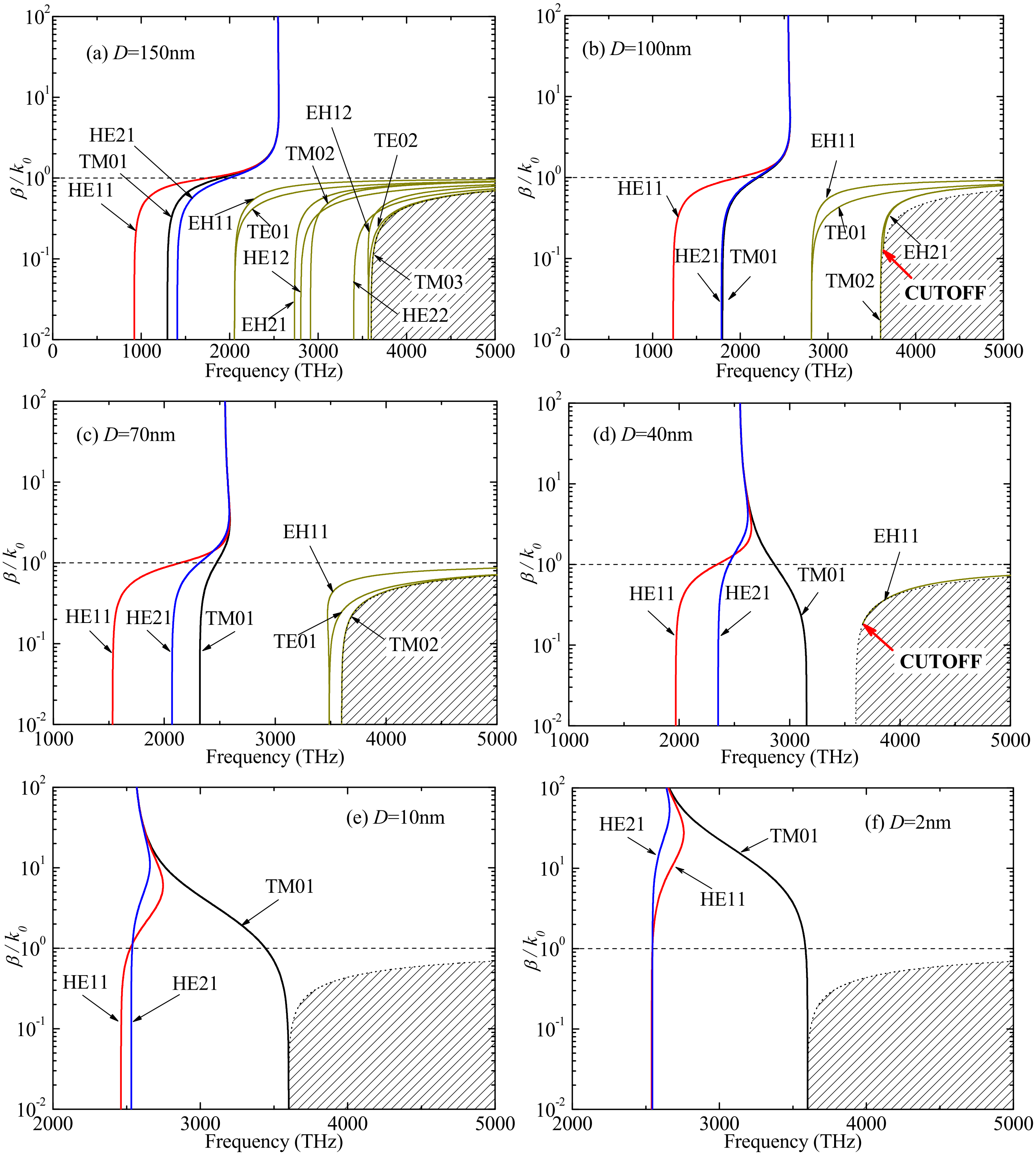

**Figure 4**

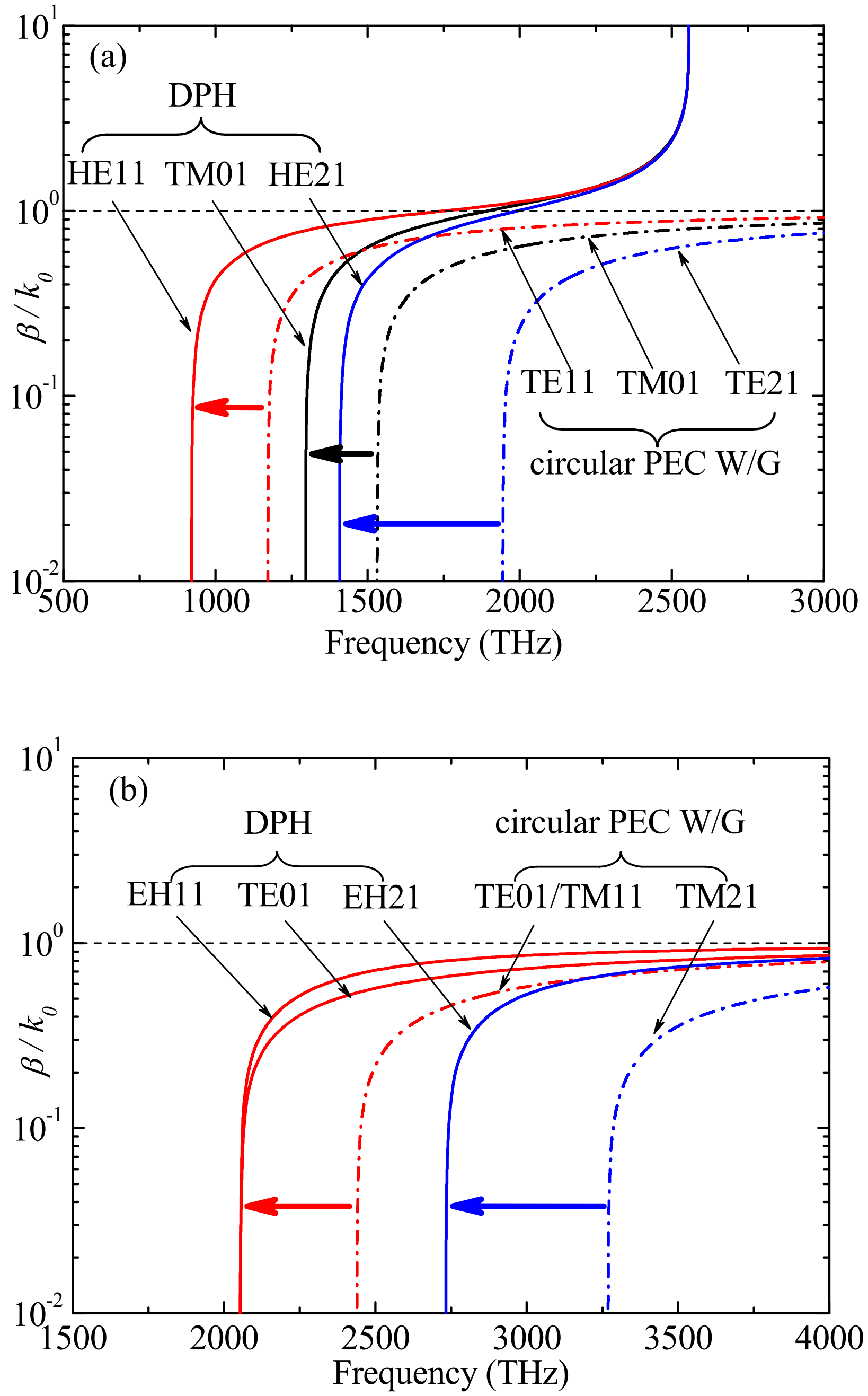

# Figure 5

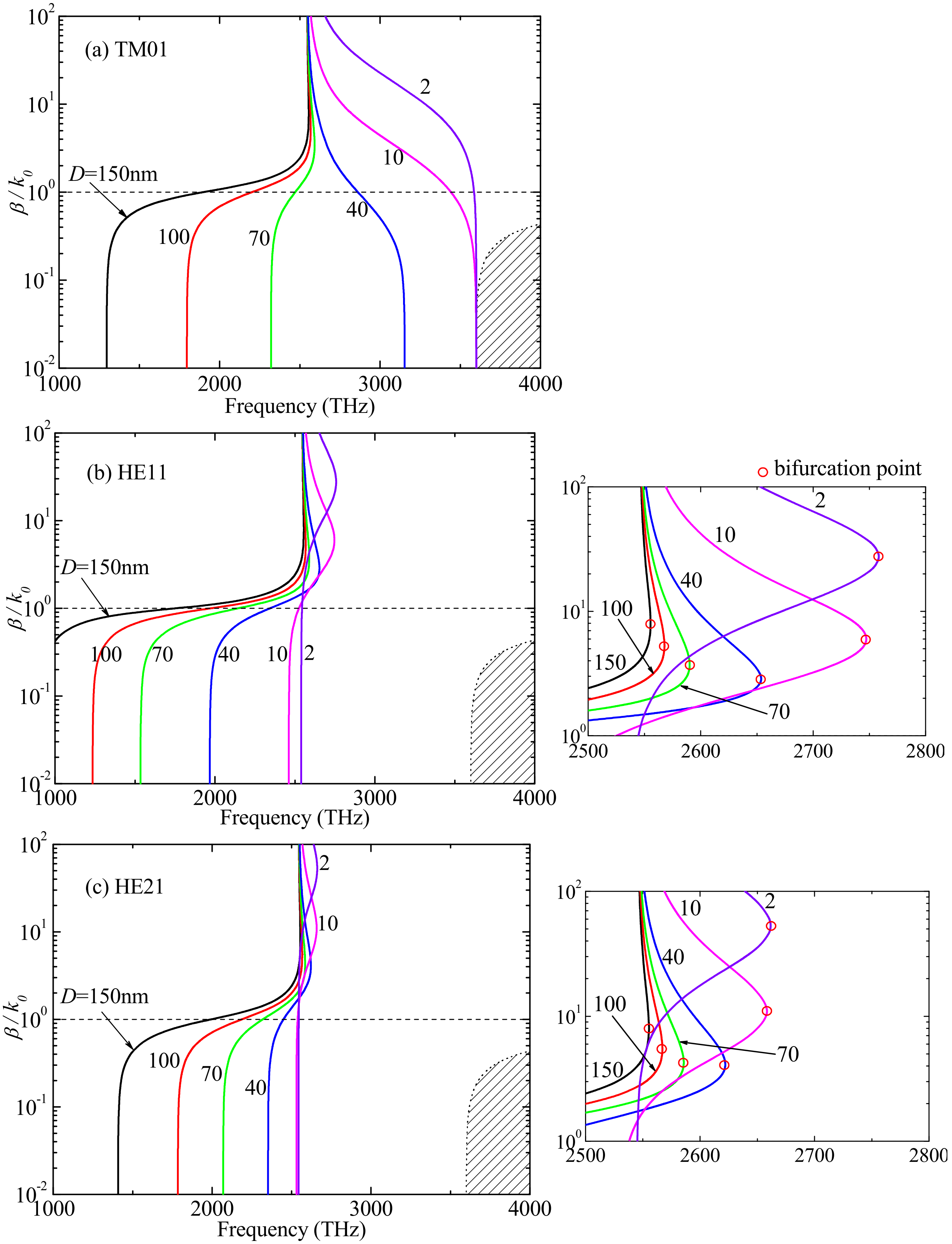

**Table 1.** Axial field components, transverse propagation constants, and primary and secondary characteristic functions included in characteristic equation (2) for DPHs.

| | **CWG mode** $\left(\mu_{r1}\varepsilon_{r1} > \bar{\beta}^2 (> \mu_{r2}\varepsilon_{r2})\right)$ | **SPP mode** $\left(\bar{\beta}^2 > \mu_{r1}\varepsilon_{r1}\right)$ |
|---|---|---|
| Axial Field Components | $E_{z1} = A_m J_m(k_1 r)$<br>$H_{z1} = B_m J_m(k_1 r)$<br>$E_{z2} = C_m K_m(k_2 r)$<br>$H_{z2} = D_m K_m(k_2 r)$ | $E_{z1} = A_m I_m(k_1 r)$<br>$H_{z1} = B_m I_m(k_1 r)$<br>$E_{z2} = C_m K_m(k_2 r)$<br>$H_{z2} = D_m K_m(k_2 r)$ |
| Transverse Propagation Constants | $k_1 = k_0\left(\mu_{r1}\varepsilon_{r1} - \bar{\beta}^2\right)^{1/2}$<br>$k_2 = k_0\left(\bar{\beta}^2 - \mu_{r2}\varepsilon_{r2}\right)^{1/2}$ | $k_1 = k_0\left(\bar{\beta}^2 - \mu_{r1}\varepsilon_{r1}\right)^{1/2}$<br>$k_2 = k_0\left(\bar{\beta}^2 - \mu_{r2}\varepsilon_{r2}\right)^{1/2}$ |
| Primary Characteristic Functions | $F_1 = \frac{\varepsilon_{r1}}{k_1}\frac{J'_m(k_1 a)}{J_m(k_1 a)} + \frac{\varepsilon_{r2}}{k_2}\frac{K'_m(k_2 a)}{K_m(k_2 a)}$<br>$F_2 = \frac{\mu_{r1}}{k_1}\frac{J'_m(k_1 a)}{J_m(k_1 a)} + \frac{\mu_{r2}}{k_2}\frac{K'_m(k_2 a)}{K_m(k_2 a)}$<br>$G = \frac{m\beta}{k_0 a}\left(\frac{1}{k_1^2} + \frac{1}{k_2^2}\right)$ | $F_1 = \frac{\varepsilon_{r1}}{k_1}\frac{I'_m(k_1 a)}{I_m(k_1 a)} - \frac{\varepsilon_{r2}}{k_2}\frac{K'_m(k_2 a)}{K_m(k_2 a)}$<br>$F_2 = \frac{\mu_{r1}}{k_1}\frac{I'_m(k_1 a)}{I_m(k_1 a)} - \frac{\mu_{r2}}{k_2}\frac{K'_m(k_2 a)}{K_m(k_2 a)}$<br>$G = \frac{m\beta}{k_0 a}\left(\frac{1}{k_1^2} - \frac{1}{k_2^2}\right)$ |
| Secondary Characteristic Functions | $P = \frac{1}{k_1 a}\left\{\frac{J_{m-1}(k_1 a)}{J_m(k_1 a)} - \frac{m}{k_1 a}\right\}$<br>$Q = \frac{1}{k_2 a}\left\{\frac{K_{m-1}(k_2 a)}{K_m(k_2 a)} + \frac{m}{k_2 a}\right\}$<br>$R = \left\{\frac{m\bar{\beta}}{a^2}\left(\frac{1}{k_1^2} + \frac{1}{k_2^2}\right)\right\}^2$ | $P = \frac{1}{k_1 a}\left\{\frac{I_{m-1}(k_1 a)}{I_m(k_1 a)} - \frac{m}{k_1 a}\right\}$<br>$Q = -\frac{1}{k_2 a}\left\{\frac{K_{m-1}(k_2 a)}{K_m(k_2 a)} + \frac{m}{k_2 a}\right\}$<br>$R = \left\{\frac{m\bar{\beta}}{a^2}\left(\frac{1}{k_1^2} - \frac{1}{k_2^2}\right)\right\}^2$ |

For the axial field components, the propagation factor $\exp\left[j(\omega t - m\theta - \beta z)\right]$ is tacitly assumed and omitted. $A_m$, $B_m$, $C_m$, and $D_m$ are constants.